\documentclass[aps,twocolumn,superscriptaddress,showpacs,pre]{revtex4}
\usepackage{graphicx}

\begin{document}
\title{Bertrand's paradox: a physical solution}

\author{P. Di Porto}
\affiliation{Dipartimento di Fisica, Universit\`{a} dell'Aquila, L'Aquila, Italy}

\author{B. Crosignani}
\affiliation{Department of Applied Physics, California Institute of Technology, Pasadena, California, 91125}

\author{A. Ciattoni}
\affiliation{Consiglio Nazionale delle Ricerche, CNR-SPIN, 67100 L'Aquila, Italy \\ and Dipartimento di Fisica, Universit\`{a} dell'Aquila, 67100
L'Aquila, Italy}

\author{H. C. Liu}
\affiliation{Department of Applied Physics, California Institute of Technology, Pasadena, California, 91125}

\date{\today}

\begin{abstract}
We present a conclusive answer to Bertrand's paradox, a long standing open issue in the basic physical interpretation of probability. The paradox
deals with the existence of mutually inconsistent results when looking for the probability that a chord, drawn at random in a circle, is longer than
the side of an inscribed equilateral triangle. We obtain a unique solution by substituting chord drawing with the throwing of a straw of {\it finite
length} $L$ on a circle of radius $R$, thus providing a satisfactory operative definition of the associated experiment. The obtained  probability
turns out to be a function of the ratio $L/R$, as intuitively expected.
\end{abstract}

 \maketitle

Bertrand's paradox is a basic example of the intrinsic ambiguity in the concept of randomness. It is associated with the probability that a chord,
drawn {\it at random} in a circle, is longer than the side of an equilateral triangle inscribed in the circle. The paradoxical nature of the problem
was originally stated by Bertrand \cite{Berta}, who showed how different solutions can be obtained based on different assumptions of equal {\it a
priori} probabilities. Three situations are typically reported in the literature (see, e.g., \cite{Mazoo}): 1) to fix one end of the chord on the
circle and draw the diameter through the fixed end: all chords lying within $\pm 30$ degrees satisfy the length condition; 2) to draw a diameter
through the midpoint of the given chord: chords intersecting the diameter between $1/4$ and $3/4$ of its length will have the required length; 3) to
choose a point anywhere within the circle and construct a chord with the chosen point as its midpoint: if the midpoint lies in a circle of radius
$R/2$, the length requirement will be again fulfilled. They lead to the result of uniform probability density $p= 1/3$, $1/2$ , $1/4$, respectively.
Although these results, as well as many other possible ones, (see, e.g., ref.\cite{Marin}) are mutually inconsistent, they are all correct. Actually,
it is the very statement of the problem that is not satisfactory since the concept of {\it drawing a chord at random} is not uniquely defined, the
random elements being not quantities but geometrical objects such as points, lines and angles which are assumed to be uniformly distributed
\cite{Marin}.

The problem has been restated by Jaynes \cite{Jayne} in a different way, adopting a more physical perspective: a long straw is tossed at random onto
a circle and the probability is sought that, given that it falls so to intercept the circle, the resulting chord is longer than the side of the
inscribed equilateral triangle. By imposing the requirement of invariant probability density, he was able to show that the only solution compatible
with the tossing of a long straw is $p=1/2$, corresponding to case 2) above. However, his procedure still exhibits a somewhat limiting feature, i.e.,
the obviously finite length $L$ of the straw is not explicitly taken into account, the straw only needing to be {\it long}. In fact, Jaynes invariant
properties imply a common final result, i.e., a probability $p$ independent from the radius $R$ of the circle. As we shall see, this is actually true
only if $L$ is much larger than $R$. In this respect, we underline that in any problem requiring absolute randomness two criteria must be fulfilled :
statistical equivalence of all relevant parameters and consideration of involved finite quantities.

In order to meet the above criteria, in this Letter we deal with a finite straw length, facing the problem in the more complete form: a straw of
length L is tossed at random onto a circle of radius R; given that it falls so that it intersects the circle, what is the probability that the chord
thus defined is longer than the side of the inscribed equilateral triangle? The relevance of our procedure is two-fold. First, we obtain a
well-defined answer more general and physically sound than that of Ref.\cite{Jayne}; second, our solution depends on the ratio $L/R$ as a priori
desirable and its limit for $L/R \gg 1$ turns out to be $1/2$, in agreement with Ref.\cite{Jayne}. We wish to underline the essential role played by
the assumption of a finite value of $L$ in devising a physically meaningful experiment in connection with Bertrand's problem.

Let us consider an horizontal surface S over which we draw a circle of radius $R$. We have at our disposal a straw of length L and assume, in order
to avoid unwanted edge effects, the linear dimensions of S to be much larger than both L and R. The straw is tossed at random onto S. What is the
meaning of {\it at random}? In this context, the only sensible answer is that all positions and orientations on S are statistically equivalent: given
a point of the straw (e.g., an extreme), its probability density of falling somewhere on S, as well as that of the straw orientation, are uniform.
Due to the large extension of S, most of times the straw will not intercept the circle: nevertheless, by repeatedly tossing the straw, a chord will
eventually be formed a number of times large enough to give meaning to Bertrand's paradox. Below, we evaluate the associated probability, conditional
to the straw intercepting the circle , and express it as a function of  L/R.

We adopt two apparently distinct approaches. In the first, we determine the probability $P$ by mainly hinging upon translational invariance; in the
second, we evaluate $P$ by essentially exploiting rotational invariance. The resulting equality of the obtained numerical values confirms the role
played by the finite length of the straw in clarifying Bertrand's paradox.

\begin{figure}
\includegraphics[width=0.5\textwidth]{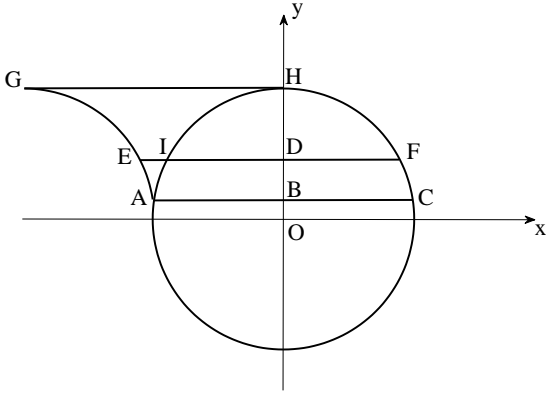}
\caption{$\overline{OH} =R$; $\overline{OB} = \sqrt{R^2-L^2/4}$; $\overline{OD} = R/2$; $\overline{AC}=\overline{EF}=\overline{GH}=L$;
$\overline{IF}=R \sqrt{3}$.}
\end{figure}

First approach - With reference to Fig.1, we assume the straw of length $L$ to have intersected the circle in a direction parallel to the x-axis, at
a distance $y$ from it. In order to evaluate the probability $P(L/R)$ of obtaining a chord longer than $R \sqrt{3}$ (we anticipate the intuitive
final dependence of $P$ on the ratio $L/R$), we consider the three intervals : a) $0<L/R\le \sqrt{3}$, b) $\sqrt{3} \le L/R \le 2$, c) $L/R \ge 2$.
In case a), since no chord can obviously exceeds $R \sqrt{3}$, we have
\begin{equation} \label{Liu1}
P(L/R)= 0, \quad  L \le R \sqrt{3}.
\end{equation}
In case b), chords can be formed provided $y \ge \sqrt{R^2-L^2/4}$ (see Fig.1). Among them, the ones longer than $R \sqrt{3}$ are contained in the
region $\sqrt{R^2-L^2/4} \le y < R/2$, while the ones shorter than $R \sqrt{3}$ pertain to the complementary region $R/2 < y < R$. On the other hand,
our randomness assumption implies that the left extreme of the chord outside the circle has a uniform probability of falling anywhere in the region
GHA having as contour the segment GH and the two circumference arches HA and AG (the last being obtained translating the arch CH by the distance L).
Thus, $P(L/R)$ is given by the ratio between the area of the sub-region EIA and that of the region GHA, i.e.,
\begin{equation} \label{Liu2}
P(L/R) = \frac{\displaystyle \int_{\sqrt{R^2-(L/2)^2}}^{R/2} dy \left[ L - 2\sqrt{R^2-y^2} \right]}{\displaystyle \int_{\sqrt{R^2-(L/2)^2}}^{R} dy
\left[ L - 2\sqrt{R^2-y^2} \right]}, \quad  R\sqrt{3} \le L \le 2R.
\end{equation}
\begin{figure}
\includegraphics[width=0.5\textwidth]{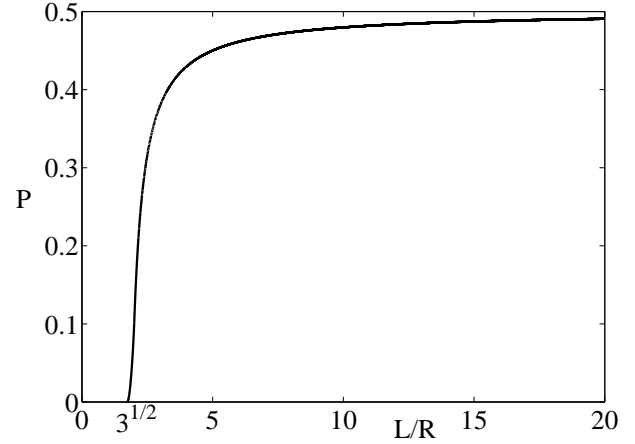}
\caption{Probability of a chord to be longer than $L \sqrt{3}$ (side of the inscribed equilateral triangle).}
\end{figure}
In case c), chords are formed for all values of $y$ between $0$ and $R$, so that
\begin{equation} \label{Liu3}
P(L/R) = \frac{\displaystyle \int_{0}^{R/2} dy \left[ L - 2\sqrt{R^2-y^2} \right]}{\displaystyle \int_{0}^{R} dy \left[ L - 2\sqrt{R^2-y^2} \right]},
\quad  L \ge 2R.
\end{equation}
After performing the integrations, Eqs.(\ref{Liu2}) and (\ref{Liu3}) respectively furnish
\begin{widetext}
\begin{equation} \label{Liu4}
P(L/R) = \frac{\frac{L}{2R} \left[ 1 -\sqrt{1-\left(\frac{L}{2R}\right)^2} \right] + \arcsin \sqrt{1-\left(\frac{L}{2R}\right)^2}
-\frac{\sqrt{3}}{4}-\frac{\pi}{6}}{\frac{L}{R} \left[ 1 - \frac{1}{2} \sqrt{1-\left(\frac{L}{2R}\right)^2} \right] + \arcsin
\sqrt{1-\left(\frac{L}{2R}\right)^2} -\frac{\pi}{2} }, \quad R\sqrt{3} \le L \le 2R,
\end{equation}
\end{widetext}
and
\begin{equation} \label{Liu5}
P(L/R) = \frac{\frac{L}{2R} - \frac{\sqrt{3}}{4} -\frac{\pi}{6}} {\frac{L}{R} - \frac{\pi}{2}}, \quad L \ge 2R.
\end{equation}
The  plot of $P(L/R)$ for all values of $L/R$ corresponding to Eqs. \ref{Liu1}, \ref{Liu4} and \ref{Liu5}, is reported in Fig.2. As expected, for
$L/R>>1$ we recover, as a particular case, Jaynes' result \cite{Jayne}, that is $P(L/R) =1/2$.

\begin{figure}
\includegraphics[width=0.5\textwidth]{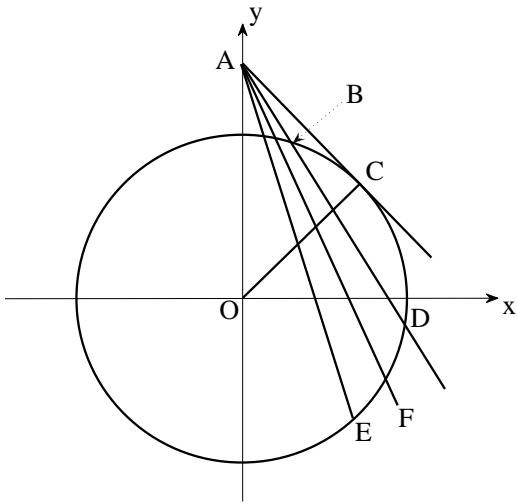}
\caption{$\overline{OA}=r$; $\overline{OC}=R$; $\widehat{OAC} = \theta_t(r)$; $\overline{BD} = R \sqrt{3}$; $\widehat{OAD} = \theta_e(r)$;
$\overline{AE}=L$; $\widehat{OAE}=\theta_L(r)$; $\overline{AF}=L$ (generic straw position); $\widehat{OAF} = \theta$.}
\end{figure}

Second approach - With reference to Fig.3, let us identify the tangent AC to the circle drawn from A, the associated angle $\theta_t (r) =
\widehat{OAC}$ ($ \tan \theta_t (r) = R/\sqrt{r^2-R^2}$), and the angle $\theta_e (r) = \widehat{OAD}$ such that $\overline{BD} = R \sqrt{3}$ ($ \tan
\theta_e (r) = R/\sqrt{4R^2-r^2}$, as easily shown by simple geometry). If we wish $A$ and $\theta$ to be the position of an extreme and the
orientation of our straw, respectively, such that a chord is formed, two necessary conditions need to be satisfied, that is $\overline{AG} < L$
(implying $r < r_m \equiv \sqrt{L^2+R^2}$) and  $\theta < \theta_t(r)$. If we consider (ignoring the case $L < R \sqrt{3}$  for which Eq.(\ref{Liu1})
obviously holds) the case $R \sqrt{3} \le L \le 2R$, there is, for every $r<r_m$, a value $\theta_L(r)$ of $\theta$ under which no chord is formed;
it can be easily checked, looking at Fig.3 and using the cosine rule, $\theta_L(r)$ to obey the relation $\cos \left[ \theta_L(r) \right] =
(r^2+L^2-R^2)/2rL$. Chords longer than $R \sqrt{3}$ can be formed only if $\theta_e(r) > \theta_L(r)$: thus, the value $r_c$ of $r$, such that no
chord longer than $R\sqrt{3}$ exists for $r>r_c$, has to fulfill the relation $\cos\left[\theta_e(r)\right]= \cos\left[\theta_L(r)\right]$, which
yields $r_c=\sqrt{L^2+R^2-LR\sqrt{3}}$.

The above considerations, and the assumed statistical equivalence of all positions A and orientations $\theta$ yield
\begin{equation} \label{Pb1}
P(L/R) = \frac{\displaystyle \int_R^{\sqrt{L^2+R^2-LR\sqrt{3}}} dr \: r \left[ \theta_e(r) - \theta_L(r)\right]} {\displaystyle
\int_R^{\sqrt{L^2+R^2}} dr \: r \left[ \theta_t(r) - \theta_L(r)\right]},   \quad R\sqrt{3} \le L \le 2R.
\end{equation}
In the case $L \ge 2R$, then, for $L-R <r<r_m$, $\theta_L$ is again the limit under which there is no chord, while for $R<r<L-r$ this limit vanishes.
Accordingly
\begin{widetext}
\begin{equation} \label{Pb2}
P(L/R) = \frac{\displaystyle \int_{L-R}^{\sqrt{L^2+R^2-LR\sqrt{3}}} dr \: r \left[ \theta_e(r) - \theta_L(r)\right] + \int_R^{L-R} dr \: r
\theta_e(r)} {\displaystyle \int_{L-R}^{\sqrt{L^2+R^2}} dr \: r \left[ \theta_t(r) - \theta_L(r)\right] + \int_R^{L-R} dr \: r \theta_t(r)},   \quad
L \ge 2R.
\end{equation}
\end{widetext}
Equations (\ref{Pb1}) and (\ref{Pb2}) can be interpreted as follows. Consider the 3-dimensional space $\Sigma$ of points $Q\equiv[x,y,\theta]$, where
$x$ and $y$ label the coordinates of the extreme A and $\theta$ the straw orientation on the surface S. Equations (\ref{Pb1}) and (\ref{Pb2})
represent the ratio between the $\Sigma$-volume spanned by the points Q for which the straw forms a chord larger than $R\sqrt{3}$ and the
$\Sigma$-volume spanned by the totality of the points Q for which the straw forms a chord of any possible length. The integrals in Eqs.((\ref{Pb1})
and (\ref{Pb2}) can be analytically performed and their numerical evaluation exactly reproduces the results of the first approach as plotted in
Fig.(2). We wish to note that the equivalence of the two methods (the first mainly hinging upon translational invariance and the second on rotational
invariance) is a simple consequence of the fact that they describe the same physical experiment. On the contrary, the standard solutions $p=1/2$ and
p=$1/3$ \cite{Mazoo,Marin}, respectively refer to a rather artificial drawing of chords parallel to a given direction or chords of different
orientation originating from a fixed point on the circumference.

In conclusion, we believe that our approach provides the natural solution to Bertrand's paradox. In fact, allowing the position of a fixed point of a
straw of finite length and the straw orientation to be uniformly distributed is the physical implementation of complete randomness. Experiments to
test the validity of the numerical behavior reported in Fig.(2) can be easily implemented by throwing a straw of length $L$ on a large surface on
which a circle of radius $R$ has been drawn, or by throwing a ring of radius $R$ on a large surface on which a segment of length $L$ has been drawn.


\begin{thebibliography} {aa}
\bibitem{Berta} J. Bertrand, {\it Calcul des Probabilit\'{e}s} (Gauthier-Villars, Paris, 1889).
\bibitem{Mazoo} R. B. Mazo, {\it Brownian motion: Fluctuations, Dynamics and Applications} (Clarendon Press, Oxford, 2002).
                N. G. Van Kampen, {\it Stochastic Processes in Physics and Chemistry} (Elsevier, 2007).
\bibitem{Marin} L. Marinoff, Philosophy of Science \textbf{61}, 1 (2004).
\bibitem{Jayne} E. T. Jaynes, Foundations of Physics \textbf{3}, 477 (1973).
\end{thebibliography}
\end{document}